\documentstyle[hip-artc]{article}
\volnumber{7}  \edyear{1998}  \frompage{000} \topage{000}                

\title{Correlator of topological charge densities at low $Q^2$
 in QCD: connection with proton spin problem.}
\authors{\twerm B. L. Ioffe\\
{\normalsize
\hspace*{-8pt} Institute of Theoretical  and Experimental Physics, \\
\hspace*{-8pt} 117218, Moscow, B.Cheremushkinskaya 25,  Russia,\\
\hspace*{-8pt} e-mail: ioffe@vitep5.itep.ru }}

\abstract{
Vacuum correlator of topological charge densities $\chi(q^2)$  at low $q^2$
in QCD is discussed: its value $\chi(0)$  and first derivative
$\chi^{\prime}(0)$ at $q^2=0$  and the contributions of pseudoscalar
quasi-Goldstone bosons to $\chi(q^2)$  at low $q^2$. The QCD sum
rule \cite{1}, giving the connection of $\chi^{\prime}(0)$ (for massless
quarks)  with the part of the proton spin $\Sigma$, carried by $u,d,s$
quarks, is presented.  From the requirement of selfconsistancy of the sum
rule the values of $\chi^{\prime}(0)$ and $\Sigma$ were found. The same
value of $\chi^{\prime}(0)$ follows also from the experimental data on
$\Sigma$. The contributions of $\pi$  and $\eta$ to $\chi(q^2)$ are
calculated basing on low energy theorems. In such calculation the $\pi-\eta$
mixing, expressed in terms of quark mass ratios is of importance.}

\begin{document}

\maketitle

\def\la{\mathrel{\mathpalette\fun <}}
\def\ga{\mathrel{\mathpalette\fun >}}
\def\fun#1#2{\lower3.6pt\vbox{\baselineskip0pt\lineskip.9pt
\ialign{$\mathsurround=0pt#1\hfil##\hfil$\crcr#2\crcr\sim\crcr}}}

\newcommand{\be}{\begin{equation}}
\newcommand{\ee}{\end{equation}}

\section{Introduction.}

The  existence of topological quantum number is a very specific feature of
non-abelian quantum field theories and, particularly, QCD. Therefore, the
study of properties of the topological charge density operator in QCD

\be
Q_5(x)= \frac{\alpha_s}{8\pi} G^n_{\mu \nu}(x)\tilde{G}^n_{\mu\nu}(x)
\label{1}
\ee
and of the corresponding vacuum correlator

\be
\chi(q^2) = i\int d^4x e^{iqx} \langle 0\mid T\left \{ Q_5(x),~Q_5(0)\right
\}\mid 0 \rangle
\label{2}
\ee
is of a great theoretical interest. (Here $G^n_{\mu\nu}$ is gluonic field
strength tensor, $\tilde{G}_{\mu\nu}=$
$(1/2)\varepsilon_{\mu\nu\lambda\sigma} G_{\lambda \sigma}$ is its dual, $n$
are
the colour indeces, $n=1,2,... N^2_c-1$, $N_c$ is the number of colours,
$N_c=3$ in QCD). The existence of topological quantum numbers in non-abelian
field theories was first discovered by Belavin et al.\cite{2}, their connection
with non-conservation of $U(1)$ chirality was established by t'Hooft
\cite{3}.
Gribov \cite{4}  was the first, who understood, that instanton
configurations in Minkowski space realize the tunneling transitions between
states with different topological numbers.
Crewther \cite{5}  derived Ward identities related to $\chi(0)$,
which allowed him to prove the theorem, that $\chi(0)=0$  in any theory
where it is at least one massless quark. An important step in the
investigation of the properties of $\chi(q^2)$ was achieved by Veneziano
\cite{6} and Di Vecchia and Veneziano \cite{7}. These authors considered the
limit $ ~N_c\to \infty$. Assuming that in the theory there are $N_f$ light
quarks with the masses $m_i \ll M$, where $M$ is the characteristic scale of
strong interaction, Di Vecchia and Veneziano found that

\be
\chi(0) = \langle 0\mid \bar{q}q\mid 0\rangle \Biggl (
\sum^{N_f}_i~\frac{1}{m_i}\Biggr )^{-1},
\label{3}
\ee
where $\langle 0\mid \bar{q}q\mid 0\rangle $ is the common value of quark
condensate for all light quarks and the terms of the order $m_i/M$ are
neglected. \footnote{The definition of $\chi(q^2)$ used  above
in eq.(\ref{2}), differs by sign from the definition used in
\cite{4}-\cite{6}.} The concept of $\theta$-term in the Lagrangian was
succesfully exploited in \cite{6} in deriving of (\ref{3}). Using the same
concept and studying the properties of the Dirac operator Leutwyler and
Smilga \cite{8}  succeeded in proving eq.3 at any $N_c$ for the case of two
light quarks, $u$ and $d$.  How this method can be generalized to the case
of $N_f=3$  is explained in the recent paper \cite{9}.

In this paper I discuss $\chi(q^2)$  in the domain of low $\mid q^2 \mid$,
i.e. I suppose that $q^2/M^2$ is a small
parameter and restrict myself to the terms linear in this ratio. In this
domain must be accounted terms $\chi(0)$, $\chi^{\prime}(0)q^2$  as well as
the contributions of low mass pseudoscalar quasi--Goldstone bosons $\pi$ and
$\eta$, resulting to nonlinear $q^2$ dependence.

Strictly speaking, only nonperturbative part of $\chi(q^2)$ has definite
meaning. The perturbative part is divergent and its contribution depends on
the renormalization procedure. For this reason only nonperturbative part of
$\chi(q^2)$ (\ref{2}) with perturbative part subtracted will be considered
here.  In ref.1 on the basis of QCD sum rules in the external fields the
connection of $\chi^{\prime}(0)$ (its nonperturbative part)  with the part
of the proton spin $\Sigma$, carried by $u,d,s$  quarks, was established.
From the requirement of the selfconsistancy of the sum rule it was obtained

\be
\chi^{\prime}(0) = (2.4 \pm 0.6) \times 10^{-3}~GeV^2
\label{4}
\ee
in the limit of massless $u,d$  and $s$  quarks. The close to (\ref{4})
value follows also by the use of experimental data on $\Sigma$. At low $q^2$
the terms, proportional to quark masses, are related to the contributions of
light pseudoscalar mesons as intermediate states in the
the correlator (\ref{2}). These
contributions are calculated below.  In such calculation for the case of
three quarks the mixing of $\pi^0$ and $\eta$  is of importance and it is
accounted.

The presentation of the material in the paper is the following.
In Sec.2 the QCD sum rule for $\Sigma$  is derived. The important
contribution to the sum rule comes from $\chi^{\prime}(0)$. This allows to
determine $\chi^{\prime}(0)$ in two ways: using the experimental data on
$\Sigma$ and from the requirement of the selfconsistancy of the sum rule.
Both methods give the same value of $\chi^{\prime}(0)$, given by eq.\ref{4}.
In Sec.3 the
low energy theorems related to $\chi(0)$  are rederived with the account of
possible anomalous equal-time commutator terms. (In \cite{5}-\cite{7} it was
implicitly assumed that these terms are zero). In Sec.4 the case of one and
two light quarks are considered. It is proved, that the mentioned above
commutator terms are zero indeed and for the case of two quarks eq.3 is
reproduced without using $N_c \to \infty$ limit and the concepts of
$\theta$--terms.  In   Sec.5 the case of three $u,d,s$ light quarks is
considered in the approximation $m_u,m_d \ll m_s$. The problem of mixing of
$\pi^0$  and $\eta$  states \cite{10,11} is formulated and corresponding
 formulae are
presented. The account of $\pi-\eta$ mixing allows one to get eq.(\ref{3})
from low energy theorems, formulated in Sec.2 for the case of three light
quarks.  (At $m_u, m_d \ll m_s$ it coincides with the two quark case). In
Sec.5 the $q^2$--dependence of $\chi(q^2)$ was found at low $\mid q^2 \mid$
in the leading nonvanishing order in $q^2/ M^2$ as well as in $m_q/M$.

\section{QCD sum rule for $\Sigma$. Connection of $\Sigma$ and
$\chi^{\prime}(0)$.}

As well known, the parts of the nucleon spin carried by
$u, d$
and $s$-quarks are determined from the measurements of the first moment
of spin dependent nucleon structure function $g_1(x, Q^2)$

\begin{equation}
\Gamma_{p,n}(Q^2) = \int \limits_{0}^{1} dx g_{1;p,n} (x, Q^2)
\label{5}
\end{equation}
The data allows one to find the value of $\Sigma$ -- the part of
nucleon spin carried by three flavours of light quarks
$\Sigma = \Delta u + \Delta d + \Delta s$,
where $\Delta u, \Delta d, \Delta s$ are the parts of nucleon spin
carried by $u,d,s$ quarks. On the basis of the operator product expansion
(OPE) $\Sigma$ is related to the proton matrix element of the flavour
singlet axial current $j^0_{\mu 5}$

\begin{equation}
2 ms_{\mu} \Sigma = \langle p, s \vert j^0_{\mu 5} \vert p,s \rangle,
\label{6}
\end{equation}
where $s_{\mu}$ is the proton spin 4-vector, $m$ is the proton mass.
The renormalization scheme in the calculation of perturbative QCD
corrections to $\Gamma_{p,n}$ can be arranged in such a way that
$\Sigma$ is scale independent.

An attempt to calculate $\Sigma$ using QCD sum rules in external fields
was done in ref.\cite{12}. Let us shortly recall the idea. The polarization
operator

\begin{equation}
\Pi(p) = i \int d^4 x e^{ipx} \langle 0 \vert T \{ \eta(x),
\bar{\eta} (0) \} \vert 0 \rangle
\label{7}
\end{equation}
was considered, where
\begin{equation}
\eta(x) = \varepsilon^{abc} \Biggl (u^a(x) C \gamma_{\mu} u ^b(x)
\Biggr ) \gamma _{\mu} \gamma_5 d^c(x)
\label{8}
\end{equation}
is the current with proton quantum numbers \cite{13}, $u^a, d^b$ are quark
fields, $a,b,c$ are colour indeces. It is assumed that the term
\begin{equation}
\Delta L = j^0_{\mu 5} A_{\mu}
\label{9}
\end{equation}
where $A_{\mu}$ is a constant singlet axial field, is added to QCD Lagrangian.
In the weak axial field approximation $\Pi(p)$ has the form

\begin{equation}
\Pi(p) = \Pi^{(0)} (p) + \Pi^{(1)}_{\mu} (p) A_{\mu}.
\label{10}
\end{equation}
$\Pi^{(1)}_{\mu}(p)$ is calculated in QCD by
OPE at $p^2 < 0, \vert
p^2 \vert \gg R^{-2}_c$, where $R_c$ is the confinement radius. On the
other hand,  using dispersion relation, $\Pi^{(1)}_{\mu} (p)$ is
represented by the contribution of the physical states, the lowest of
which is the proton state. The contribution of excited states is
approximated as a continuum and suppressed by the Borel transformation. The
desired answer is obtained by equalling of these two representations. This
procedure can be applied to any Lorenz structure of $\Pi^{(1)}_{\mu} (p)$ ,
but as was argued in \cite{14},\cite{15} the best accuracy can be obtained by
considering the chirality conserving structure $2 p_{\mu} \hat{p} \gamma_5$.

An essential ingredient of the method is the appearance of induced by
the external field vacuum expectation values (v.e.v). The most
important of them in the problem at hand is

\begin{equation}
\langle 0 \vert j^0_{\mu 5} \vert 0 \rangle_A \equiv 3 f^2_0 A_{\mu}
\label{11}
\end{equation}
of dimension 3. The constant $f^2_0$ is related to QCD topological
susceptibility. Using (\ref{9}), we can write
$$
\langle 0 \vert j^0_{\mu 5} \vert 0 \rangle_A = lim_{q \to 0} ~ i \int~
d^4 xe^{iqx} \langle 0 \vert T \{ j^0_{\nu 5} (x), j^0_{\mu 5} (0) \}
\vert 0 \rangle A_{\nu} \equiv
$$
\begin{equation}
\equiv lim_{q \to 0} P_{\mu \nu} (q) A_{\nu}
\label{12}
\end{equation}
The general structure of $P_{\mu \nu} (q)$ is

\begin{equation}
P_{\mu \nu} (q) = -P_L(q^2) \delta_{\mu \nu} + P_T(q^2) (-\delta_{\mu
\nu} q^2 + q_{\mu} q_{\nu})
\label{13}
\end{equation}
Because of anomaly there are no massless states in the spectrum of the
singlet polarization operator $P_{\mu \nu}$ even for massless quarks.
$P_{T,L}(q^2)$ also have no kinematical singularities at $q^2 = 0$ .
Therefore, the nonvanishing value $P_{\mu \nu} (0)$ comes entirely from
$P_L(q^2)$. Multiplying $P_{\mu \nu} (q)$ by $q_{\mu} q_{\nu}$, in the
limit of massless $u, d, s$ quarks we get

$$
q_{\mu} q_{\nu} P_{\mu \nu} (q) = -P_L (q^2) q^2 = N^2_f
(\alpha_s / 4 \pi)^2 i \int~ d^4 xe^{iqx} \times
$$
\begin{equation}
\times \langle 0 \vert T {G^n_{\mu \nu} (x) \tilde{G}^n_{\mu \nu} (x),
G^m_{\lambda \sigma} (0) \tilde{G}^m _{\lambda \sigma} (0)} \vert 0 \rangle,
\label{14}
\end{equation}
where $G^n_{\mu \nu}$ is the gluonic field strength, $\tilde{G}_{\mu
\nu} = (1/2) \varepsilon_{\mu \nu \lambda \sigma} G_{\lambda \sigma}$.(The anomaly
condition was used, $N_f = 3$.).
Going to the limit $q^2 \to 0$, we have

\begin{equation}
f^2_0 = -(1/3) P_L(0) = \frac{4}{3} N^2_f \chi^{\prime} (0),
\label{15}
\end{equation}
where $\chi(q^2)$ is defined by (\ref{2}).

According to Crewther theorem \cite{5}), $\chi(0) = 0$ if there is
at least one massless quark. The attempt to find $\chi^{\prime}(0)$ itself
by QCD sum rules failed: it was found \cite{12} that OPE does not converge
in the domain of characteristic scales for this problem. However, it
was possible to derive the sum rule, expressing $\Sigma$ in terms of
$f^2_0$ (\ref{11}) or $\chi^{\prime}(0)$. The OPE up to dimension $d=7$ was
performed in ref.\cite{12}. Among the induced by the external field v.e.v.'s
besides (\ref{11}), the v.e.v. of the dimension 5 operator

\begin{equation}
g \langle 0 \vert \sum \limits_{q} \bar{q} \gamma_{\alpha} (1/2) \lambda
^n \tilde{G}^n_{\alpha \beta} q \vert 0 \rangle _A \equiv 3 h_0
A_{\beta}, ~~~ q = u, d, s
\label{18}
\end{equation}
was accounted and the constant $h_0$ was estimated using a special sum
rule,\\
$h_0 \approx 3 \times 10^{-4} GeV^4$ . There were also accounted the
gluonic condensate $d = 4$ and the square of quark condensate $d= 6$
(both times the external $A_{\mu}$ field operator, $d=1$). However, the
accuracy of the calculation was not good enough for reliable
calculation of $\Sigma$ in terms of $f^2_0$: the necessary requirement
of the method -- the weak dependence of the result on the Borel
parameter was not well satisfied.

In \cite{1} the accuracy of the calculation was improved by going to
higher order terms in OPE up to dimension 9 operators. Under the
assumption of factorization -- the saturation of the product of
four-quark operators by the contribution of an intermediate vacuum
state -- the dimension 8 v.e.v.'s are accounted (times $A_{\mu}$):

\begin{equation}
-g \langle 0 \vert \bar{q} \sigma_{\alpha \beta} (1/2) \lambda^n
G^n_{\alpha \beta} q \cdot \bar{q} q \vert 0 \rangle = m^2_0 \langle 0
\vert \bar{q} q \vert 0 \rangle^2,
\label{19}
\end{equation}
where $m^2_0 = 0.8 \pm 0.2 ~GeV^2$
was determined in \cite{16}.
In the framework of the same factorization hypothesis the induced by
the external field v.e.v. of dimension 9

\begin{equation}
\alpha_s \langle 0 \vert j^{(0)}_{\mu 5} \vert 0 \rangle_A \langle 0
\vert \bar{q} q \vert 0 \rangle^2
\label{20}
\end{equation}
is also accounted. In the calculation  the following expression for
the quark Green function in the constant external axial field was used
\cite{15}:
$$ \langle 0 \vert T \{ q^a_{\alpha}(x),~ \bar{q}^b_{\beta}
(0) \} \vert 0 \rangle_A = i \delta^{ab} \hat{x}_{\alpha \beta} / 2 \pi^2
x^4 + $$
$$ + (1/2 \pi^2) \delta^{ab} (A x) (\gamma_5 \hat{x})_{\alpha
\beta} / x^4 - (1/12) \delta^{ab} \delta_{\alpha \beta} \langle 0 \vert
\bar{q} q \vert 0 \rangle + $$
$$ +(1/72)i\delta^{ab} \langle 0 \vert
\bar{q} q \vert 0 \rangle (\hat{x} \hat{A} \gamma_5 - \hat{A} \hat{x}
\gamma_5)_{\alpha \beta} +
$$
\begin{equation} + (1/12) f^2_0 \delta^{ab}
(\hat{A} \gamma_5)_{\alpha \beta} + (1/216) \delta^{ab} h_0 \Biggl [(5/2)
  x^2 \hat{A} \gamma_5 - (A x) \hat{x} \gamma_5 \Biggr ]_{\alpha \beta}
\label{21}
\end{equation}
The terms of the third power  in $x$-expansion of
quark propagator proportional to $A_{\mu}$ are omitted in (\ref{21}),
 because they do not contribute to the tensor structure of $\Pi_{\mu}$ of
 interest.  Quarks are considered to be in the constant external gluonic
 field and quark and gluon QCD equations of motion are exploited (the
 related formulae are given in \cite{17}). There is also an another source
 of v.e.v. $h_0$  to appear besides the $x$-expansion of quark propagator
 given in eq.(\ref{21}): the quarks in the condensate absorb the soft
 gluonic field emitted by other quark. A similar situation takes place also
 in the calculation of the v.e.v. (\ref{20}) contribution. The accounted
 diagrams with dimension 9 operators have no loop integrations.  There are
 others v.e.v.  of dimensions $d \leq 9$ particularly containing gluonic
 fields.  All of them, however, correspond to at least one loop integration
 and are suppressed by the numerical factor $(2 \pi)^{-2}$. For this reason
 they are disregarded.

The sum rule for $\Sigma$ is given by
$$
\Sigma + C_0 M^2 = -1 + \frac{8}{9 \tilde{\lambda}^2_N} e^{m^2/M^2}
\left \{a^2 L^{4/9} +
6\pi^2 f^2_0 M^4 E_1 \Biggl (\frac{W^2}{M^2} \Biggr ) L^{-4/9} + \right.
$$
\begin{equation}
\left.   +
14 \pi^2 h_0 M^2 E_0 \Biggl (\frac {W^2}{M^2} \Biggr ) L^{-8/9} -
\frac{1}{4}~ \frac{a^2 m^2_0}{M^2} - \frac{1}{9} \pi \alpha_s f^2_0~
\frac{a^2}{M^2} \right \}
\label{22}
\end{equation}

Here $M^2$ is the Borel parameter, $\tilde{\lambda}_N$ is defined as
$\tilde{\lambda}^2_N = 32 \pi^4 \lambda^2_N = 2.1 ~ GeV^6$,
$\langle 0 \vert \eta \vert p \rangle = \lambda_N v_p,$
where $v_p$ is proton spinor, $W^2$ is the continuum threshold, $W^2 =
2.5 ~GeV^2$,

\begin{equation}
a = -(2 \pi)^2 \langle 0 \vert \bar{q} q \vert 0 \rangle = 0.55 ~ GeV^3
\label{23}
\end{equation}
$$
E_0(x) = 1 - e^{-x} , ~~~ E_1(x) = 1 - (1 + x)e^{-x}
$$
$L = ln (M/\Lambda)/ ln (\mu/\Lambda),~~~
\Lambda = \Lambda_{QCD} = 200 ~ MeV$ and the normalization point  $\mu$
was chosen $\mu = 1 ~ GeV$. When deriving (\ref{22}) the sum rule for the
nucleon mass was exploited what results in  appearance of the first
term, -1, in the right hand side (rhs) of (\ref{22}). This term absorbs the contributions of
the bare loop, gluonic condensate as well as $\alpha_s$ corrections
to them and essential part of terms, proportional to $a^2$ and $m^2_0 a^2$.
The values of the parameters, $a, \tilde{\lambda}^2_N, W^2$ taken
above were chosen by the best fit of the sum rules for the nucleon mass
(see \cite{18}, Appendix B) performed at $\Lambda = 200 ~ MeV$. It can be
shown, using the value of the ratio $2m_s/ (m_u + m_d) = 24.4 \pm 1.5$
\cite{19} that $a(1~ GeV) = 0.55 ~ GeV^3$ corresponds to $m_s(1 ~ GeV) = 153
~ MeV$.  $\alpha_s$ corrections are accounted in the leading order (LO) what
results in appearance of anomalous dimensions. Therefore $\Lambda$ has the
meaning of effective $\Lambda$ in LO. The unknown constant $C_0$ in the
left-hand side (lhs) of (\ref{22}) corresponds to the contribution of inelastic
transitions $p \to N^* \to interaction~ with A_{\mu} \to p$ (and in inverse
order). It cannot be determined theoretically and may be found from $M^2$
dependence of the rhs of (\ref{22}) (for details see  \cite{18,20}). The
necessary condition of the validity of the sum rule is $\vert \Sigma \vert
\gg \vert C_0 M^2 \vert exp [(-W^2 + m^2)/M^2]$ at characteristic values of
$M^2$ \cite{20}. The contribution of the last term in the rhs of (\ref{22}))
is negligible.  The sum rule (\ref{22}) as well as the sum rule for the
nucleon mass is reliable in the interval of the Borel parameter $M^2$  where
the last term of OPE is small, less than $10-15\%$ of the total and the
contribution of continuum does not exceed $40-50\%$. This fixes the
interval $0.85 < M^2 < 1.4 ~ GeV^2$.The $M^2$-dependence of the rhs of
(\ref{22})
at $f^2_0 = 3 \times 10^{-2}~ GeV^2$ is plotted in fig.1a. The complicated
expression in rhs of (\ref{22}) is indeed an almost linear function of $M^2$
in the given interval!  This fact strongly supports the reliability of the
approach. The best values of $\Sigma = \Sigma^{fit}$ and $C_0 = C^{fit}_0$
are found from the $\chi^2$ fitting procedure

\begin{equation}
\chi^2 = \frac{1}{n} \sum \limits ^{n}_{i=1}~ [\Sigma^{fit} -
C^{fit}_0 M^2_i - R(M^2_i)]^2 = min,
\label{24}
\end{equation}
where $R(^2)$  is the rhs of (\ref{22}).

\begin{figure}[htb]
\vspace*{-1.0cm}
                 \insertplot{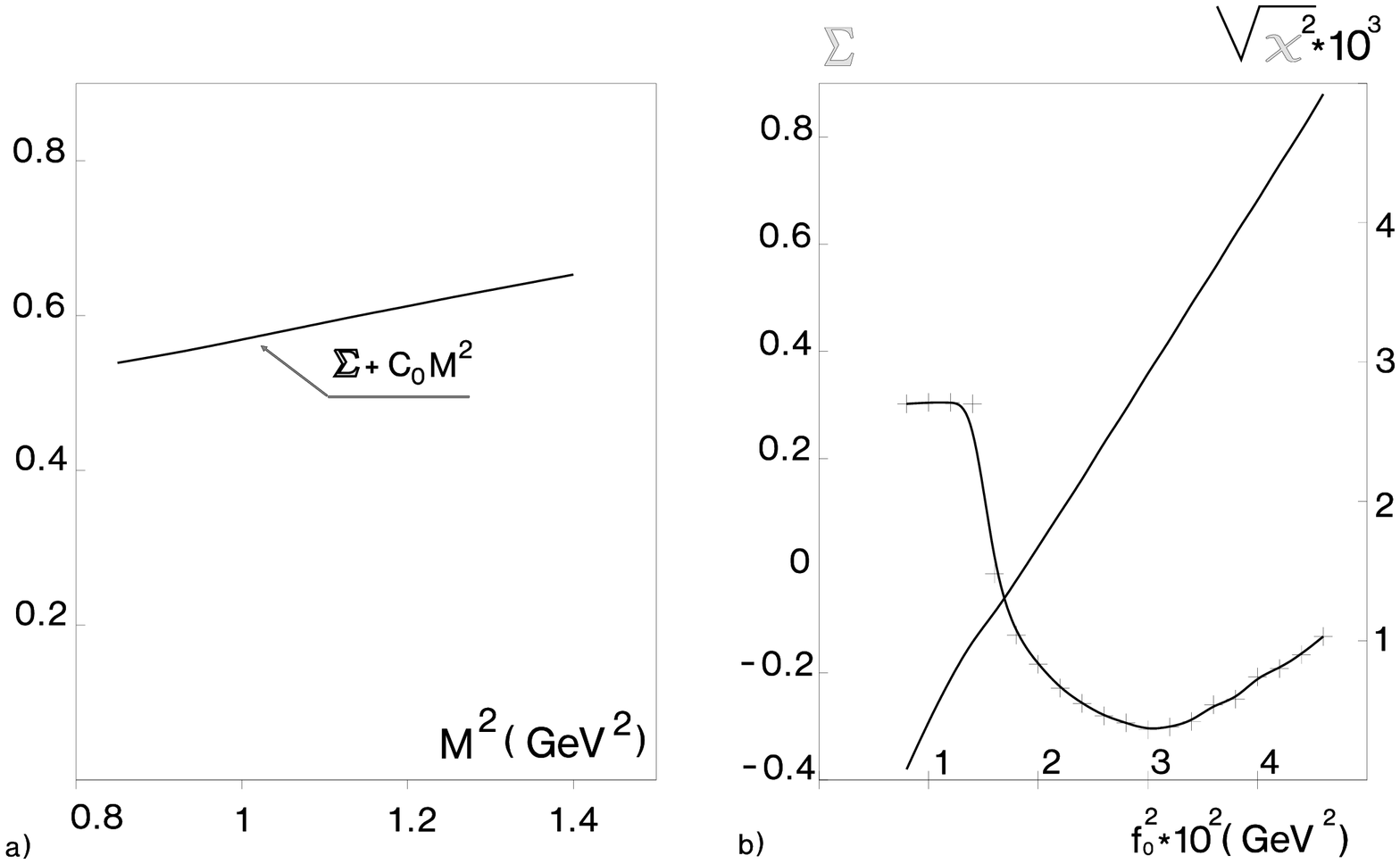}
\vspace*{-1.0cm}
\caption[]{ a) The $M^2$-dependence of $\Sigma + C_0 M^2$ at $f^2_0 =
3 \times 10^{-2}~GeV^2$; b) $\Sigma$  (solid line, left ordinate
axis) and $\sqrt{\chi^2}$, (dashed line, right ordinate axis).
as a functions of $f^2_0$. }
\label{fig1}
\end{figure}

The values of $\Sigma$ as a function of $f^2_0$  are
plotted in fig.1b together with $\sqrt{\chi^2}$.
In our approach the gluonic
contribution cannot be separated and is included in $\Sigma$. The
experimental value of $\Sigma$ can be estimated \cite{21,22} (for discussion
see \cite{23}) as $\Sigma = 0.3 \pm 0.1$. Then from fig.1b we have $f^2_0 =
(2.8 \pm 0.7) \times 10^{-2}~GeV^2$ and $\chi^{\prime}(0) = (2.3 \pm 0.6)
 \times 10^{-3}~GeV^2$. The error in $f^2_0$ and $\chi^{\prime}$ besides
the experimentall error includes the uncertainty in the sum rule estimated
as equal to the contribution of the last term in OPE (two last terms in
Eq.(\ref{22})
and a possible role of NLO $\alpha_s$ corrections.
Allowing the deviation of  $\sqrt{\chi^2}$ by a factor  1.5 from the minimum
we get $\chi^{\prime}= (24\pm 0.6)\times 10^{-3} GeV^2$ and
$\Sigma =0.32\pm 0.17$ from the requirement of selfconsistency of the sum
rule.
At $f^2_0 = 2.8 \times 10^{-2}~ GeV^2$  the value of the constant $C_0$
found from the fit is $C_0 = 0.19 ~ GeV^{-2}$.  Therefore, the mentioned
above necessary condition of the sum rule validity is well satisfied.

Let us discuss the role of various terms of OPE in the sum rules (\ref{22}).
To analyze it the  sum rule (\ref{22}) was considered for 4 different cases,
i.e.  when it is taken into consideration:  a) only  contribution of the
operators  up to d=3 (the term --1 and the term, proportional to $f^2_0$ in
(\ref{22})); b) contribution of the operators up to d=5 (the term $\sim h_0$
is added); c) contribution of the operators up to d=7 (three first terms in
(\ref{22})), d) the result (\ref{22}), i.e. all operators up to d=9.  For
this analysis the value of $f^2_0 = 0.03 ~GeV^2$ was chosen, but the
conclusion appears to be the same for all more or less reasonable choice of
$f^2_0$.  Results of the fit of the sum rules are shown in  Table 1 for all
four cases. The fit is done in the region of Borel masses $0.9 <M^2 < 1.3~
GeV^2$. In the first column the values of $\Sigma$ are shown , in the second
- values of the parameter C, and in the third - the ratio $\gamma = \vert
\sqrt{\chi^2}/\Sigma \vert$, which is the real parameter, describing
reliability of the fit. From  the table one can see, that reliability of the
fit monotonously improves with increasing of the number of accounted terms
of OPE and is quite satisfactory in the case $d$

\begin{table}[h]
\caption[]

\begin{center}
\begin{tabular}{|c|c|c|c|}
\hline
case & $\Sigma$ & $C(GeV^{-2})$ & $\gamma$\\ \hline
a) & -0.019 & 0.31 & $10^{-1}$\\ \hline
b) & 0.031 & 0.3 & $5.10^{-2}$\\ \hline
c) & 0.54 & 0.094 & $9.10^{-3}$\\ \hline
d) & 0.36 & 0.21 & $1.3 \cdot 10^{-3}$ \\ \hline
\end{tabular}

\end{center}

\end{table}


\section{Low energy theorems.}

Consider QCD with $N_f$  light quarks, $m_i \ll M \sim 1~GeV$, $i=1,...
N_f$. Define the singlet (in flavour) axial current by

\be
j_{\mu 5}(x) = \sum^{N_f}_i \bar{q}_i(x)\gamma_{\mu}\gamma_5 q(x)
\label{25}
\ee
and the polarization operator

\be
P_{\mu\nu} (q) = i\int d^4x e^{iqx} \langle 0\mid T~ \{j_{\mu 5}(x),
j_{\mu 5}(0)~\} \mid 0 \rangle.
\label{26}
\ee
The general form of the polarization operator is:

\be
P_{\mu \nu} (q) = -P_L(q^2)\delta_{\mu\nu} + P_T (q^2)(-\delta_{\mu\nu} q^2
+ q_{\mu}q_{\nu})
\label{27}
\ee
Because of anomaly the singlet axial current is nonconserving:

\be
\partial_{\mu}j_{\mu 5}(x) = 2N_f Q_5(x) + D(x),
\label{28}
\ee
where $Q_5(x)$ is given by (1) and

\be
D(x) = 2i\sum^{N_f}_i m_i\bar{q}_i (x)\gamma_5 q_i(x)
\label{29}
\ee
It is well known, that even if some light quarks are massless, the
corresponding Goldstone bosons, arising from spontaneous violation of chiral
symmetry do not contribute to singlet axial channel (it is the solution of
$U(1)$ problem), i.e. to polarization operator $P_{\mu\nu}(q)$. $P_L(q^2)$
also have no kinematical singularities at $q^2=0$. Therefore

\be
P_{\mu\nu} (q)q_{\mu}q_{\nu} = -P_L(q^2)q^2
\label{30}
\ee
vanishes in the limit $q^2 \to 0$. Calculate the left-hand side (lhs) of
(\ref{30})
in the standard way -- put $q_{\mu}q_{\nu}$ inside the integral in
(\ref{26}) and integrate by parts. (For this it is convenient to represent
the polarization operator in the coordinate space as a function of two
coordinates $x$ and $y$.) Going to the limit $q^2 \to 0$ we have

$$
\lim_{q^2\to 0}P_{\mu\nu} (q)q_{\mu}q_{\nu} = i\int d^4 x \langle 0 \mid T\{
2N_fQ_5(x),~2N_fQ_5(0)  $$
$$+ 2N_fQ_5(x),~D(0) + D(x), ~2N_fQ_5(0) + D(x),~D(0)\} \mid 0 \rangle $$
$$+4\sum^{N_f}_i m_i \langle \mid \bar{q}_i(0)q_i(0)\mid 0 \rangle  $$
\be
+\int d^4 x \langle  0 \mid ~[j_{05}(x),~2N_f Q_5 (0)~] \mid 0 \rangle
\delta(x_0)= 0
\label{31}
\ee
In the calculation of (\ref{31}) the anomaly condition (\ref{28}) was used.
The terms, proportional to quark condensates arise from equal time
commutator $[j_{05}(x)$, $D(0)]_{x_0=0}$, calculated by standard commutation
relations.  Relation (\ref{31}) up to the last term was first obtained by
Crewther \cite{5}.  The last term, equal to zero according to standard
commutation relations and omitted in \cite{5}-\cite{7}, is keeped. The
reason is, that we deal with very subtle situation, related to anomaly,
where nonstandard Schwinger terms in commutation relations may appear. (It
can be shown, that, in general the only Schwinger term in this problem is
given by the last term in the lhs of (\ref{31}): no others can arise.)
Consider also the correlator:

\be
P_{\mu}(q) = i\int d^4x~e^{iqx} \langle 0 \mid T \{j_{\mu 5}(x),~Q_5(0)\}
\mid 0 \rangle
\label{32}
\ee
and the product $P_{\mu}(q)q_{\mu}$  in the limit $q^2\to 0$ (or $q^2$ of
order of the $m^2_{\pi}$, where $m_{\pi}$ is the mass of Goldstone boson).
The general form of $P_{\mu}(q)$ is $P_{\mu}(q)=Aq_{\mu}$. Therefore
nonvanishing values of $P_{\mu}q_{\mu}$  in the limit $q^2 \to 0$ (or of
order of quark mass $m$, if $\mid q^2 \mid \sim m^2_{\pi}$ -- this limit will
be
also intersting for us later) can arise only from Goldstone bosons
intermediate states in (\ref{32}).
Let us us estimate the corresponding matrix elements

\be
\langle 0 \mid j_{\mu 5}\mid \pi \rangle = F q_{\mu}
\label{33}
\ee
\be
\langle 0 \mid Q_5\mid \pi \rangle = F^{\prime}
\label{34}
\ee
$F$ is of order of $m$, since in the limit of massless quarks Goldstone
bosons are coupled only to nonsinglet axial current. $F^{\prime}$ is of order
of $m^2_{\pi}f_{\pi}\sim m$, where $f_{\pi}$ is the pion decay constant (not
considered to be small), since in massless quark limit, the Goldstone boson is
decoupled
 from $Q_5$ . These estimations give

\be
P_{\mu}q_{\mu} \sim \frac{q^2}{q^2 - m^2_{\pi}} m^2
\label{35}
\ee
and it is zero at $q^2 \to 0,~m^2_{\pi}\not=0$ and of order of $ m^2$ at
$q^2\sim
m^2_{\pi}$. In what follows I will restrict myself by the terms linear in
quark masses.  So, I can put $P_{\mu}(q)q_{\mu}=0$  at $q\to 0$. The
integration by parts, in the right-hand side (rhs) of (\ref{32}) gives:

$$
\lim_{q^2\to 0} P_{\mu}(q)q_{\mu} = -\int d^4x \langle 0 \mid T \{ 2N_f
Q_5(x),~ Q_5(0) + D(x),~Q_5(0)\}\mid 0 \rangle $$
\label{36}
\be
-\int d^4 x \langle 0 \mid [~j_{05}(x),~Q_5(0)~]\mid 0 \rangle \delta(x_0) =0
\ee
After the substitution of (\ref{36}) in (\ref{31}) arise the low energy
theorem:

$$i\int d^4 x \langle 0 \mid T \{ 2N_f Q_5(x),~ 2N_fQ_5(0)\} \mid 0\rangle
$$
$$-i\int d^4 x \langle 0\mid T\{ D(x),~D(0)\} \mid 0 \rangle - 4\sum^{N_f}_i
m_i \langle 0 \mid \bar{q}_i(0)q_i(0)\mid 0 \rangle $$
\be
+ i\int d^4 x \langle 0\mid  [~j^0_{05}(x),~2N_fQ_5(0)~] \mid 0\rangle
\delta(x_0) =0
\label{37}
\ee
The low energy theorem (\ref{37}), with the last term in the lhs omitted, was
found by Crewther \cite{5}.


\section{One and two light quarks.}

Consider first the case of one massless quark, $N_f=1$, $m=0$. This case can
easily be treated by introduction of $\theta$-term in the Lagrangian,

\be
\Delta L = \theta \frac{\alpha_s}{4\pi} ~G^n_{\mu\nu}\tilde{G}^n_{\mu\nu}
\label{38}
\ee
The matrix element $\langle 0 \mid Q_5 \mid n \rangle$ between any hadronic
state $\mid n \rangle$ and vacuum is proportional

\be
\int d^4x \langle 0 \mid Q_5(x) \mid n \rangle  \sim \langle  0 \mid
\frac{\partial}{\partial\theta} lnZ\mid n \rangle_{\theta=0},
\label{39}
\ee
where $Z=e^{iL}$ and $L$ is the Lagrangian. The gauge transformation of the
quark field $\psi^{\prime}\to e^{i\alpha\gamma_5 }\psi$  results to appearance
of
the term

\be
\delta L = \alpha \partial_{\mu}j_{\mu 5} =\alpha (\alpha_s/4\pi)G^n_{\mu\nu}
\tilde{G}^n_{\mu\nu}
\label{40}
\ee
in the Lagrangian. By the choice $\alpha= -\theta$ the $\theta$-term
(\ref{38}) will be killed and $(\partial/\partial \theta)lnZ=0$. Therefore,
$\chi(0)=0$ (Crewther theorem). The first term in (\ref{37}) vanishes, as
well as the second and third, since $m=0$. From (\ref{37}) we have, that
indeed the anomalous commutator vanishes

\be
\langle 0 \mid [~j_{05}(x),~Q_5(0)~]_{x_0=0} \mid 0 \rangle =0,
\label{41}
\ee
supporting the assumptions done in \cite{5}-\cite{7}.

Let us turn  now to the case of two light quarks, $u,d,N_f=2$. This is the
case of real QCD, where the strange quark is considered as a heavy. Define
the isovector axial current

\be
j^{(3)}_{\mu 5} = (\bar{u}\gamma_{\mu}\gamma_5 u -\bar{d}\gamma_{\mu}
\gamma_5 d)/\sqrt{2}
\label{42}
\ee
and its matrix element between the states of pion and vacuum

\be
\langle 0 \mid j^{(3)}_{\mu 5} \mid \pi \rangle = f_{\pi} q_{\mu},
\label{43}
\ee
where $q_{\mu}$ is pion 4-momentum, $f_{\pi}=133~MeV$. Multiply (\ref{43}) by
$q_{\mu}$. Using Dirac equations for quark fields, we have

$$\frac{2i}{\sqrt{2}}\langle  0\mid m_u
\bar{u}\gamma_{5}u -
m_d\bar{d}\gamma_{5}d \mid \pi \rangle
= \frac{i}{\sqrt{2}}
\langle  0\mid (m_u + m_d)(\bar{u}\gamma_{5}u - \bar{d}\gamma_{5}d) $$
\be
+ (m_u - m_d)((\bar{u}\gamma_{5}u + \bar{d}\gamma_{5}d)
\mid \pi \rangle = f_{\pi} m^2_{\pi},
\label{44}
\ee
where $m_u$, $m_d$  are $u$ and $d$ quark masses. The ratio of the matrix
elements in lhs  of (\ref{44}) is of order

\be
\frac{\langle 0 \mid \bar{u}\gamma_{5}u +
\bar{d}\gamma_{5}d \mid \pi \rangle }
{\langle 0 \mid \bar{u}\gamma_{5}u -
\bar{d}\gamma_{5}d \mid \pi \rangle } \sim \frac{m_u-m_d}{M},
\label{45}
\ee
since the matrix element in the numerator violates isospin and this violaton
(in
the absence of elecntomagnetism, which is assumed)  can arise from the
difference $m_u-m_d$ only. Neglecting this matrix element we have from
(\ref{44})

\be
\frac{i}{\sqrt{2}}\langle 0 \mid
\bar{u}\gamma_{5}u -
\bar{d}\gamma_{5}d
\mid \pi \rangle  = \frac{f_{\pi}m^2_{\pi}}{m_u+m_d}
\label{46}
\ee
Let us find $\chi(0)$ from low energy sum rule (\ref{37}) restricting
ourself to the terms linear in quark masses. Since $D(x)\sim m$, the only
intermediate state contributing to  the matrix element

\be
\int d^4 x \langle  0 \mid T\{D(x),~ D(0)\} \mid 0 \rangle
\label{47}
\ee
in (\ref{37}) is the one-pion state. Define

\be
D_q = 2i(m_u\bar{u}\gamma_5 u + m_d \bar{d}\gamma_5 d).
\label{48}
\ee
Then

$$\langle 0 \mid D_q \mid \pi \rangle = i\langle 0 \mid (m_u+m_d)
(\bar{u}\gamma_{5}u + \bar{d}\gamma_{5}d) + (m_u-m_d)
(\bar{u}\gamma_{5}u - \bar{d}\gamma_{5}d) \mid \pi \rangle$$

\be
= \sqrt{2} \frac{m_u-m_d}{m_u+m_d} f_{\pi}m^2_{\pi},
\label{49}
\ee
where the matrix element of singlet axial current was neglected  and
(\ref{46}) was used. The substitution of (\ref{49}) into (\ref{47}) gives

$$i\int d^4 x e^{iqx} \langle 0 \mid T \{ D_q(x),~D_q(0)\} \mid 0
\rangle_{q\to 0}= \lim_{q\to 0} \left \{ -\frac{1}{q^2-m^2_{\pi}}2\Biggl (
\frac{m_u - m_d}{m_u+m_d}\Biggr )^2 f^2_{\pi} m^2_{\pi}\right \} $$
\be
= - 4 \frac{(m_u - m_d)^2}{m_u+m_d}\langle 0 \mid \bar{q}q \mid 0 \rangle
\label{50}
\ee
In the last equality in (\ref{50}) Gell-Mann-Oakes-Renner relation \cite{24}

\be
\langle 0 \mid \bar{q}q \mid 0 \rangle =
-\frac{1}{2}~\frac{f^2_{\pi}m^2_{\pi}}{m_u+m_d}
\label{51}
\ee
was substituted as well the $SU(2)$ equalities

\be
\langle 0 \mid \bar{u}u \mid 0 \rangle =
\langle 0 \mid \bar{d}d \mid 0 \rangle \equiv
\langle 0 \mid \bar{q}q \mid 0 \rangle.
\label{52}
\ee
From (\ref{37}) and (\ref{50}) we finally get:

\be
\chi(0) = i\int d^4 x \langle 0 \mid T\{ Q_5(x),~Q_5(0) \} \mid 0 \rangle =
\frac{m_um_d}{m_u + m_d}\langle 0 \mid \bar{q}q \mid 0 \rangle
\label{53}
\ee
in concidence with eq.3.

 In a similar way matrix element
$\langle 0 \mid Q_5 \mid \pi \rangle$ can be found. Consider

\be
\langle 0 \mid j_{\mu 5} \mid \pi \rangle = F q_{\mu}
\label{54}
\ee
The estimation of $F$  gives

\be
F\sim \frac{m_u - m_d}{M} f_{\pi}
\label{55}
\ee
and after multiplying of (\ref{54}) by $q_{\mu}$ the rhs of (\ref{54}) can be
neglected.
In the lhs we have

\be
\langle 0\mid D_q\mid \pi \rangle + 2N_f \langle 0\mid Q_5\mid \pi \rangle
=0
\label{56}
\ee
The substitution of (\ref{49}) into (\ref{56}) results in

\be
\langle 0 \mid Q_5 \mid \pi \rangle = -\frac{1}{2\sqrt2}\frac{m_u -
m_d}{m_u+m_d}
f_{\pi}m^2_{\pi}
\label{57}
\ee
The relation of this type (with a wrong numerical coefficient) was found in
\cite{11}, the correct formula was presented in \cite{25}. From comparison
 of (\ref{53}) and (\ref{57}) it is clear, that it would be wrong to
 calculate $\chi(0)$ by accounting only pions as intermediate states in the
 lhs of (\ref{53}) -- the constant terms, reflecting the necessity of
 subtraction terms in dispersion relation and represented by proportional to
 quark condensate terms in (\ref{37}) are extremingly important. The
 cancellation of Goldstone bosons pole terms and these constant terms
 results in the Crewther theorem -- the vanishing of $\chi(0)$, when one of
 the quark masses, e.g. $m_u$  is going to zero.


\section{Three light quarks.}

Let us dwell on the real QCD case of three light quarks, $u,d$ and $s$. Since
the ratios $m_u/m_s$, $m_d/m_s$ are small, less than 1/20, account them only
in the leading order. When the $u$ and $d$  quark masses $m_u$ and $m_d$
are not assumed to be equal, the quasi-Goldstone states $\pi^0$  and $\eta$
are no more states of pure isospin 1 and 0 correspondingly: in both of these
states persist admixture of other isospin \cite{10,11} proporional to
$m_u-m_d$.
(The violation of isospin by electromagnetic interaction is small in the
problem under investigation \cite{10} and can be neglected. The
$\eta^{\prime} - \eta$ mixing is also neglected.) In order to treat the
problem it is convenient to introduce pure isospin 1 and 0 pseudoscalar
meson fields $\varphi_3$ and $\varphi_8$ in $SU(3)$ octet and the
corresponding states $\mid P_3 \rangle$, $\mid P_8 \rangle$ \cite{10,25}.
Then in the $SU(3)$ limit
\be
\langle 0\mid j^{(3)}_{\mu 5} \mid P_3 \rangle = f_{\pi}q_{\mu}
\label{58}
\ee
\be
\langle 0\mid j^{(8)}_{\mu 5} \mid P_8 \rangle = f_{\pi}q_{\mu},
\ee
\label{59}
where

\be
j^{(8)}_{\mu 5} = (\bar{u}\gamma_{\mu}\gamma_5 u +
\bar{d}\gamma_{\mu}\gamma_5 d - 2\bar{s}\gamma_{\mu}\gamma_5 s)/\sqrt{6}
\label{60}
\ee
and $j^{(3)}_{\mu 5}$ is given by (\ref{42}). The states $\mid P_3 \rangle$,
$\mid P_8 \rangle$ are not eigenstates of the Hamiltonian. In the free
Hamiltonian

\be
H = \frac{1}{2}\tilde{m}^2_{\pi} \varphi^2_3 +
\frac{1}{2}\tilde{m}^2_{\eta}\varphi^2_8
+ \langle P_8 \mid P_3 \rangle \varphi_3\varphi_8  + \mbox{kinetic terms}
\label{61}
\ee
the nondiagonal term $\sim \varphi_3\varphi_8$ is present (
$\tilde{m}^2_{\pi}$ and $\tilde{m}^2_{\eta}$ in (\ref{61}) coincides with
$m^2_{\pi}$ and $m^2_{\eta}$ up to terms quadratic in  $\langle P_8 \mid P_3
\rangle$).
The nondiagonal term was calculated in \cite{10} on the basis of PCAC  and
current algebra (see also \cite{11})

\be
\langle P_8 \mid P_3 \rangle = \frac{1}{\sqrt{3}}m^2_{\pi}
\frac{m_u-m_d}{m_u+m_d}
\label{62}
\ee
The physical  $\pi$ and $\eta$ states arise after orthogonalization of the
Hamiltonian (\ref{61})

$$
\hspace*{-0.8cm}\mid \pi \rangle = cos \theta \mid P_3 \rangle - sin \theta
\mid P_8
\rangle$$
\be
\mid \eta\rangle = sin \theta \mid P_3 \rangle + cos \theta \mid P_8 \rangle
\label{63}
\ee
where the mixing angle $\theta$ is given by (at small $\theta$) \cite{10,11}:

\be
\theta = \frac{\langle  P_8 \mid P_3 \rangle}{m^2_{\eta} - m^2_{\pi}}\approx
\frac{\langle P_8 \mid P_3 \rangle}{m^2_{\eta}} =
\frac{1}{\sqrt{3}}~\frac{m^2_{\pi}}{m^2_{\eta}}~\frac{m_u-m_d}{m_u+m_d}
\label{64}
\ee
In terms of the fields $\varphi_3$  and $\varphi_8$ PCAC relations take the
form \cite{10};

\be
\partial_{\mu}j^{(3)}_{\mu 5} = f_{\pi}(m^2_{\pi}\varphi_3 +
\langle  P_8 \mid P_3 \rangle \varphi_8)
\label{65}
\ee
\be
\partial_{\mu}j^{(8)}_{\mu 5} = f_{\pi}(m^2_{\pi}\varphi_8 +
\langle  P_8 \mid P_3 \rangle \varphi_3)
\label{66}
\ee
Our goal now is to calculate the contribution of pseudoscalar octet states
to the second term in the lhs of (\ref{37}). It is convenient to use the
full set of orthogonal states $\mid P_3 \rangle$, $\mid P_8 \rangle $ as the
basis.  Use the notation

\be
D = D_q + D_s ~~~~~~~~D_s = 2im_s\bar{s}\gamma_5 s,
\label{67}
\ee
where $D_q$  is given by (\ref{48}). The matrix element

\be
\langle 0 \mid D_q \mid P_3 \rangle = \sqrt{2} \frac{m_u-m_d}{m_u + m_d}
f_{\pi}m^2_{\pi}
\label{68}
\ee
can be found by the same argumentation, as that was used in the derivation
of (\ref{49}). In order to find $\langle 0 \mid D_q \mid P_8\rangle$ take
the matrix element of eq.\ref{65} between vacuum and $\mid P_8\rangle$

\be
\langle 0 \mid \partial_{\mu}j^{(3)}_{\mu 5} \mid P_3 \rangle =
f_{\pi} \langle  P_8 \mid P_3 \rangle
\label{69}
\ee
The substitution in the lhs of (\ref{69}) of the expression for
$\partial_{\mu}j^{(3)}_{\mu 5} $ through quark fields gives

$$\langle 0 \mid \bar{u}\gamma_5 u - \bar{d}\gamma_5 d \mid P_8 \rangle = -
\frac{m_u - m_d}{m_u + m_d}
\langle 0 \mid \bar{u}\gamma_5 u + \bar{d}\gamma_5 d \mid P_8 \rangle -$$
\be
- i\sqrt{\frac{2}{3}} f_{\pi}m^2_{\pi} \frac{m_u - m_d}{(m_u + m_d)^2}
\label{70}
\ee
In a similar way take matrix element of eq. (\ref{66}) between $\langle 0
\mid$ and $\mid P_8 \rangle$  and substitute into it (\ref{70}). We get

$$\frac{i}{\sqrt{6}}\langle 0 \mid \Biggl [ m_u+m_d-\frac{(m_u
- m_d)^2} {m_u + m_d}\Biggr ] (\bar{u}\gamma_5 u + \bar{d}\gamma_5 d) - 4m_s
\bar{s}\gamma_5 s \mid P_8 \rangle$$
\be
=f^2_{\pi}m^2_{\eta} - \frac{1}{3}m^2_{\pi}f_{\pi}\Biggl (
\frac{m_u - m_d}{m_u + m_d}\Biggr )^2
\label{71}
\ee
As  follows from  $SU(3)$ symmetry of strong interaction

\be
\langle 0 \mid \bar{u}\gamma_5 u + \bar{d}\gamma_5 d \mid P_8 \rangle = -
\langle 0 \mid\bar{s}\gamma_5 s \mid P_8 \rangle
\label{72}
\ee
up to terms of order $m_q/M$, which are neglected. From
(\ref{71}),(\ref{72}) we find:

\be
\langle 0 \mid D_s\mid P_8 \rangle  = -\sqrt{\frac{3}{2}}f_{\pi}m^2_{\eta}
\Biggl [ 1 - \frac{1}{4} \frac{(m_u - m_d)^2}{m_s(m_u + m_d)}\Biggr ]
\Biggl [ 1 + \frac{m_u m_d}{m_s(m_u + m_d)}\Biggr ]
\label{73}
\ee

\be
\langle 0 \mid D_q\mid P_8 \rangle  = 4\sqrt{\frac{2}{3}}f_{\pi}m^2_{\pi}
\frac{m_u m_d}{(m_u + m_d)^2}
\label{74}
\ee
in notation (\ref{48}),(\ref{67}). When deriving (\ref{73}) the $SU(3)$
relation

\be
m^2_{\eta} = \frac{4}{3} m^2_{\pi} \frac{m_s}{m_u+m_d}
\Biggl ( 1 - \frac{1}{4} \frac{m_u + m_d}{m_s}\Biggr )
\label{75}
\ee
was used. In (\ref{73}) the small terms $\sim m_u/m_s$, $m_d/m_s$ are
accounted, because they are multiplyed by large factor $m^2_{\eta}$. In
(\ref{74})
small terms are disregarded. The matrix element $\langle 0 \mid D_s\mid P_3
\rangle $ can be found from (\ref{66}). We have

\be
\frac{1}{\sqrt{6}}
\langle 0 \mid D_q - 2D_s \mid P_3 \rangle  = f_{\pi}
\langle P_8 \mid  P_3 \rangle
\label{76}
\ee
The substitution of (\ref{68}) and (\ref{62}) into (\ref{76}) gives

\be
\langle 0 \mid D_s \mid P_3 \rangle  = 0
\label{77}
\ee
Equations (\ref{68}),(\ref{73}),(\ref{74})  and (\ref{77}) allow one to
calculate the interesting for us correlator

\be
i \int d^4 x \langle 0 \mid T \{ D(x),~D(0) \} \mid 0 \rangle
\label{78}
\ee
when the sets $\mid P_3 \rangle \langle P_3 \mid$  and
$\mid P_8 \rangle \langle P_8 \mid$ are taken as intermediate states. But
$\mid P_3 \rangle,~ \mid  P_8  \rangle$ are not the eigenstates  of the
Hamiltonian, they mix in accord with (\ref{61}). Therefore the transitions
$\langle  P_8 \mid P_3 \rangle$ arising from the mixing term in (\ref{61})
must be also accounted. There are two such terms. The one corresponds to the
transition $\langle 0\mid  D_s \mid P_8 \rangle$
$\langle P_3 \mid  D_q \mid 0 \rangle$ and its contribution to (\ref{68}) is
given by

$$\lim_{q^2\to 0} \left \{ -2 \langle 0\mid  D_s \mid P_8 \rangle
\frac{1}{q^2 - m^2_{\eta}}\langle P_8 \mid P_3 \rangle
\frac{1}{q^2 - m^2_{\pi}} \langle P_3\mid  D_q \mid 0 \rangle \right \} =$$
\be
= 2f^2_{\pi} m^2_{\pi} \Biggl ( \frac{m_u - m_d}{m_u+m_d}\Biggr )^2
\label{79}
\ee
The other corresponds to the transition between two $D_s$ operators,
where $\langle P_3 \mid P_3 \rangle$ enter as intermediate state. This
contribution is equal to:

$$\lim_{q^2\to 0} \left \{ - \langle 0\mid  D_s \mid P_8 \rangle
\frac{1}{q^2 - m^2_{\eta}}\langle P_8 \mid P_3 \rangle
\frac{1}{q^2 - m^2_{\pi}} \langle P_8 \mid P_3 \rangle
\frac{1}{q^2 - m^2_{\eta}}
\langle P_8\mid  D_s \mid 0 \rangle \right \} =$$
\be
= \frac{1}{2}f^2_{\pi} m^2_{\pi} \Biggl (\frac{m_u - m_d}{m_u+m_d}\Biggr )^2
\label{80}
\ee
It is enough to account only matrix elements, with $D_s$ operators, since
they are enhanced by large factor $m^2_{\eta}$. All others are small in the
ratio $m^2_{\pi}/m^2_{\eta}$.

Collecting all together, we get:

$$i\int d^4 x \langle 0 \mid T \{ D(x),~D(0) \} \mid 0 \rangle=
f^2_{\pi} m^2_{\pi} \left \{ 2 \Biggl ( \frac{m_u -m_d}{m_u +m_d} \Biggr
)^2 - 8 \frac{m_u m_d}{(m_u +m_d)^2} \right. $$
$$ + 2\frac{m_s}{m_u +m_d}\Biggl ( 1 + \frac{1}{4}\frac{m_u +m_d}{m_s}
\Biggr ) \Biggl [ 1 - \frac{1}{2}\frac{m_u -m_d}{m_s(m_u+ m_d)} \Biggr ]
\Biggl [ 1 - 2\frac{m_u m_d}{m_s(m_u+ m_d)} \Biggr ]$$
\be
\left. + 2  \Biggl ( \frac{m_u -m_d}{m_u +m_d} \Biggr )^2 +
\frac{1}{2}\Biggl ( \frac{m_u -m_d}{m_u+ m_d} \Biggr )^2 \right \}=
f^2_{\pi} m^2_{\pi} \Biggl [  \frac{2m_s}{m_u +m_d}
+ \frac{9}{2}\Biggl ( \frac{m_u -m_d}{m_u+ m_d} \Biggr )^2 -
\frac{5}{2}\Biggr ]
\label{81}
\ee
The first term in the figure bracket in (\ref{81}) comes from
$\langle 0\mid \ D_q\mid P_3\rangle^2$, the second -- from
$\langle 0\mid D_q\mid P_8\rangle \times \langle P_8 \mid D_s\mid
0 \rangle$, the third -- from $\langle 0\mid D_s\mid P_8\rangle^2$, the
last two terms are from (\ref{79}),(\ref{80}).  Adding to (\ref{81}) the
proportional to quark condensate term

\be
4(m_u + m_d + m_s)\langle 0 \mid \bar{q}q \mid 0 \rangle
\label{82}
\ee
in (\ref{37}), we finally get for 3 quarks at $m_u, m_d \ll m_s$

\be
i\int d^4 x \langle 0 \mid T \{ 2N_f Q_5(x),~ 2N_f Q_5(x)\} \mid 0 \rangle=
36 \frac{m_u m_d}{m_u +m_d} \langle 0 \mid \bar{q}q \mid 0 \rangle
\label{83}
\ee
and

\be
\chi(0) = \frac{m_u m_d}{m_u +m_d} \langle 0 \mid \bar{q}q \mid 0 \rangle,
\label{84}
\ee
since in this case $4 N_f^2=36$. Eq.\ref{84} coincides with eq.\ref{3},
obtained \cite{7} in $N_c \to \infty~$ limit.  This fact demonstrates, that
$N_c \to \infty~$ limit is irrelevant for determination of $\chi(0)$ (at
least for the cases of two or three light quarks and at $m_u, m_d \ll m_s$).
$\chi(0)$ for three light quarks at $m_u,m_d \ll m_s$ -- eq.\ref{84}
coincides with $\chi(0)$ in the two light quark case, [see \cite{8} and
(\ref{53})] i.e in this problem, when $m_u,m_d \ll m_s$, there is no
difference if $s$-quark is considered as a heavy or light -- it softly
appears in the theory.

Determine the matrix elements $\langle 0 \mid Q_5\mid \eta \rangle$  and
$\langle 0 \mid Q_5\mid \pi\rangle$.  Following \cite{26} consider

\be
\langle 0 \mid j_{\mu 5} \mid \eta \rangle = \tilde{F}q_{\mu}
\label{85}
\ee
$\tilde{F}$ is of order of $f_{\pi}(m_s/M)$ and  can be put to zero in our
approxamation. By taking the divergence from (\ref{85}), we have

\be
\langle 0 \mid D_s + 6Q_5 \mid \eta \rangle = 0
\label{86}
\ee
The use of (\ref{83})  gives (the $\pi-\eta$ mixing as well terms of order
$m_u/m_s,~m_d/m_s$ may be neglected here):

\be
\langle 0 \mid Q_5 \mid \eta \rangle =  \frac{1}{2}
\sqrt{\frac{1}{6}}f_{\pi}m^2_{\eta}
\label{87}
\ee
Relation (\ref{87}) was found in \cite{26}. By the same reasoning it is easy
to prove that

\be
\langle 0 \mid D_q \mid P_3\rangle + \langle 0 \mid D_s \mid P_3\rangle +
6 \langle 0 \mid Q_5 \mid P_3\rangle =0
\label{88}
\ee
The first term in (\ref{88})  is given by (\ref{68}), the second one is zero
according (\ref{77}).  For the last term we can write using (\ref{63})

\be
\langle 0 \mid Q_5 \mid P_3\rangle = \langle 0 \mid Q_5 \mid \pi \rangle +
\theta \langle 0 \mid Q_5 \mid P_8\rangle
\label{89}
\ee
Eq.'s (\ref{88}),(\ref{89}) give

\be
\langle 0 \mid Q_5 \mid \pi \rangle = -\frac{1}{2\sqrt{2}} f_{\pi}m^2_{\pi}
\frac{m_u -m_d}{m_u + m_d}
\label{90}
\ee

-- the same formula as in the case of two light quarks.

It is clear, that the presented above considerations can be generalized to the
case,
 when u,d and s-quark masses are comparable. The calculation became more
cumbersome,
but nothing principially new arises on this case.


\section{$q^2$--dependence of $\chi(q^2)$ at low $q^2$.}

Let us dwell on the  calculation of the $q^2$--dependence of $\chi(q^2)$ at
low $\mid q^2 \mid$ in QCD, restricting ourselves by the first order terms
in the ratio $q^2/M^2$, where $M$ is the characteristic hadronic scale, $M^2
\sim 1~GeV^2$.
In this domain of $q^2~\chi (q^2)$
can be represented as

\be
\chi(q^2) = \chi(0) + \chi^{\prime}(0)q^2 + R(q^2) - R(0)
\label{91}
\ee
$\chi(0)$ for the QCD case--three light quarks with  $m_u,m_d \ll m_s$ --
was determined in Sec.V. $\chi^{\prime}(0)$ (its nonperturbative part) for
massless quarks was found in \cite{1}  basing on connection of
$\chi^{\prime}(0)$ with the part of proton spin $\sum$ carried by $u,d,s$
quarks. Its numerical value is given by (\ref{4}). What is left, is the
contribution of light pseudoscalar quasi--Goldstone bosons $R(q^2)$, which
has nontrivial $q^2$--dependence and must be accounted separately. $R(q^2)$
vanishes for massless quarks and did not contribute to $\chi^{\prime}(0)$,
calculated in \cite{1}. $R(0)$ must be subtracted from $R(q^2)$ since it was
already accounted in $\chi(0)$. $R(q^2)$  can be written as
\be
R(q^2) = - \langle 0 \mid Q_5 \mid \pi \rangle^2  \frac{1}{q^2
-m^2_{\pi}} - \langle 0 \mid Q_5 \mid \eta \rangle^2  \frac{1}{q^2
-m^2_{\eta}}
\label{92}
\ee
The problem of $\eta - \pi$ mixing is irrelevant in the difference $R(q^2) -
R(0)$  in any domain $\mid q^2 \mid \sim m_{\pi}^2$ and $\mid q^2 \mid \sim
m^2_{\eta}$. The matrix elements entering (\ref{92})  are given by
(\ref{87}),(\ref{90}).  Taking the difference $R(q^2) - R(0)$ and using the
Euclidean variable $Q^2 =-q^2$, we have

\be
\chi(Q^2) = \chi(0) - \chi^{\prime}(0)Q^2 - \frac{1}{8}f^2_{\pi}Q^2 \Biggl [
\Biggl ( \frac{m_u -m_d}{m_u + m_d} \Biggr )^2
\frac{m^2_{\pi}}{Q^2+m^2_{\pi}} + \frac{1}{3}\frac{m^2_{\eta}}{Q^2
+m^2_{\eta}}\Biggr ]
\label{93}
\ee
Eq.(\ref{93}) is our final result, where $\chi(0)$ is given by (\ref{84}) and
$\chi^{\prime}(0)$ by (\ref{4}). The  accuracy of (\ref{93}) is given by the
parameters $Q^2/M^2$, $m^2_{\pi}/M^2$, $m^2_{\eta}/M^2 \ll 1$, (two last
characterize the accuracy of $SU(3)\times SU(3)$). At $Q^2\approx
m^2_{\eta}$  the last term comprise about 20\% of the second (the first term
is very small, $\chi(0)\approx -4\cdot 10^{-5}~GeV^4$). Evidently, the last
term is much bigger in the Minkovski domain, $Q^2 < 0$, since there are pion
and $\eta$  poles. As was mentioned above, $\chi(0)$ found here concides
with $\chi(0)$ obtained in \cite{7} by considering large $N_c~$ limit.
However, the $q^2$-dependence is completely different.  Namely, for
$\chi(q^2)$ in \cite{7} was found the relation (eq.($A4^{\prime}$) in
\cite{7})
\be
\chi(q^2) = - \frac{aF^2_{\pi}}{2N_c} \Biggl [ 1 - \frac{a}{N_c}\sum_i \frac{1}
{q^2 - \mu_i^2} \Biggr ]^{-1},
\label{94}
\ee
where the Goldstone boson masses $\mu_i^2$ are related to quark condensate by
\be
\mu_i^2 = -2m_i \frac{1}{F^2_{\pi}} \langle 0 \mid \bar{q} q \mid 0 \rangle
\label{95}
\ee
and $a$ is some constant of order of hadronic mass square. At $q^2=0$ follows
eq.\ref{3} for
$\chi(0)$ if the inequality $a/{N_c\mu_i^2}>>1$ is assumed. However, at
$|q^2|\sim
m^2_{\eta}$ , $m^2_{\pi}$ (\ref{94}) strongly differs from (\ref{93}): in
(\ref{94}) there are zeros at the points $q^2 = m^2_{\eta}$ , $m^2_{\pi}$,
but not poles, as it should be and as it take place in (\ref{93}). And also
the most important at low $Q^2$ hadronic term $\chi^{\prime}(0)Q^2$ is
absent in (\ref{94}).


\section{Summary.}

The $q^2$-dependence of topological charge density correlator $\chi(q^2)$ (2)
in QCD
was considered in the domain of low $q^2$.

The QCD sum rule, connecting $\chi^{\prime}(0)$  (for massless $u,d,s$
quarks) with the part of the proton spin $\Sigma$, carried by $u,d,s$
quarks in the polarized proton, is derived. The numerical value of
$\chi^{\prime}(0)$  was found from the requirement of selfconsistancy of the
sum rule. In the limit of errors the same value of $\chi^{\prime}(0)$
follows also from the experimental data on $\Sigma$.

For the cases of two and three light
quarks
the values of  $\chi(0)$, obtained earlier \cite{5}-\cite{8} were rederived
basing on
the low energy theorems and accounting of quasi-Goldstone boson
($\pi$,$\:\eta$)
contributions. No large $N_c$ limit was used and it was no appeal to the
$\theta$-dependence of QCD Lagrangian (except of the proof of absence of
anomalous
commutator in the sum rule (\ref{37})).
The only concept, which was used, was the absence of Goldstone boson
contribution as
intermediate state in the singlet (in flavour) axial current correlator in the
limit of
massless quarks.
In the three light quark case -- the case of real QCD -- the mixing of $\pi$
and $\eta$
is of importance and was widely exploited. The $q^2$ dependence of $\chi(q^2)$
was found as arising from two sources:
\begin{itemize}

\item{1}. The contribution of hadronic states (besides $\pi$ and $\eta$),
determined from the connection of $\chi^{\prime}(0)$ with the part of the
proton spin, carried by quarks.
\item{2}. The contributions of $\pi$ and
$\eta$ intermediate states. These contributions were calculated by using low
 energy theorems only.  The final result is presented in eq.\ref{93}.
\end{itemize}


\vskip 10pt
\noindent \large {\bf Acknowledgement}\normalsize \vskip 10pt
\noindent

I am very indebted to
J.Speth for the hospitality at the Institut f\"ur Kernphysik, FZ J\"ulich,
where this work was finished, and to A.v.Humboldt Foundation for financial
support of this visit.
This work was supported in part by CRDF grant
RP2-132,  Schweizerischer National Fonds grant 7SUPJ048716 and RFBR grant
97-02-16131.

\vfill\eject


\end{document}